\documentclass[letterpaper]{article} 
\usepackage{aaai2026}  
\usepackage{times}  
\usepackage{helvet}  
\usepackage{courier}  
\usepackage[hyphens]{url}  
\usepackage{graphicx} 
\usepackage{natbib}  
\usepackage{caption} 
\usepackage{algorithm}
\usepackage{algorithmic}

\usepackage{amsmath}
\usepackage{amsfonts}
\usepackage{dsfont}
\usepackage{graphicx}
\usepackage{textcomp}
\usepackage{xcolor}
\usepackage{booktabs}
\usepackage{multirow}
\usepackage{multicol}
\usepackage{microtype}
\usepackage{graphicx}
\usepackage{pdflscape}
\usepackage{fancyhdr}
\usepackage{comment}
\usepackage{makecell}
\usepackage{caption}
\usepackage{subcaption}
\usepackage{dsfont}
\usepackage{enumitem}
\usepackage{placeins}
\usepackage{enumitem}

\def\methodname{SPINRec}
\def\method{\methodname~}

\def\xu{\mathbf{x}_u}

\def\xmm1{\mathbf{x}^{1-m}}

\def\r{\mathbf{r}}

\def\v{\mathbf{v}}

\def\m{\mathbf{m}}

\def\mstar{\mathbf{m^{*}}}
\def\x{\mathbf{x}}

\def\z{\mathbf{z}}
\def\v1{\mathbf{1}}

\def\f{f_{\theta}}

\newcommand{\rank}[2]{\operatorname{rank}_{#1}^{#2}}

\def\xuremoved{\mathbf{x}_u^{\setminus K_e}}
\def\xuretained{\mathbf{x}_u^{K_e}}

\def\rankremoved{\rank{\f}{y}(\xuremoved)}

\usepackage{newfloat}
\usepackage{listings}
\DeclareCaptionStyle{ruled}{labelfont=normalfont,labelsep=colon,strut=off} 
\floatstyle{ruled}
\newfloat{listing}{tb}{lst}{}
\floatname{listing}{Listing}
\title{Fidelity-Aware Recommendation Explanations via Stochastic Path Integration}
\author{
    Oren Barkan\textsuperscript{\rm 1}\thanks{Equal contribution.}, 
    Yahlly Schein\textsuperscript{\rm 2}\footnotemark[1], Yehonatan Elisha\textsuperscript{\rm 2},\\ Veronika Bogina\textsuperscript{\rm 2}, Mikhail Baklanov\textsuperscript{\rm 2}, Noam Koenigstein\textsuperscript{\rm 2}\thanks{Corresponding author.}    
}
\affiliations{
    \textsuperscript{\rm 1}The Open University, Israel\\
    \textsuperscript{\rm 2}Tel Aviv University, Israel\\    
}

\begin{document}

\maketitle

\begin{abstract}
Explanation fidelity, which measures how accurately an explanation reflects a model’s true reasoning, remains critically underexplored in recommender systems. We introduce \method (Stochastic Path Integration for Neural Recommender Explanations), a model-agnostic approach that adapts path-integration techniques to the sparse and implicit nature of recommendation data. To overcome the limitations of prior methods, \method employs stochastic baseline sampling: instead of integrating from a fixed or unrealistic baseline, it samples multiple plausible user profiles from the empirical data distribution and selects the most faithful attribution path. This design captures the influence of both observed and unobserved interactions, yielding more stable and personalized explanations. We conduct the most comprehensive fidelity evaluation to date across three models (MF, VAE, NCF), three datasets (ML1M, Yahoo! Music, Pinterest), and a suite of counterfactual metrics, including AUC-based perturbation curves and fixed-length diagnostics. \method consistently outperforms all baselines, establishing a new benchmark for faithful explainability in recommendation. Code and evaluation tools are publicly available at {\color{blue} \url{https://github.com/DeltaLabTLV/SPINRec}}.
\end{abstract}

\section{Introduction}
\label{sec:intro}

Recent advances in recommender systems over the past decade~\cite{he2017neural,kang2018self,he2020lightgcn,barkan2019cb2cf,barkan2020neural,barkan2021cold,katz2022learning} have increasingly shaped personalized decisions across e-commerce, social media, and streaming platforms, making transparency and trust more essential than ever.~\citep{fan2022comprehensive}. Explainability in these systems is critical not only for user satisfaction but also for accountability, compliance with regulations, and user control.  However, while explainable recommendation research is rapidly expanding~\citep{zhang2020explainable,varasteh2024comparative}, most existing work focuses on user-centric aspects such as persuasiveness, clarity, or satisfaction~\citep{kunkel2019let,tintarevmeasuring}. A critical yet underexplored dimension is \emph{fidelity}, which measures how accurately explanations reflect a recommender’s actual decision process. Without fidelity, explanations may appear plausible while failing to reveal the true reasoning behind recommendations~\citep{koenigstein2025without}.



We introduce \method (Stochastic Path Integration for Neural Recommender Explanations), the first adaptation of path-integration (PI)~\citep{sundararajan2017axiomatic} to recommender systems. Unlike prior applications of PI in vision~\cite{kapishnikov2021guided,barkan2023visual,barkan2023deep,elisha2024probabilistic,barkan2025bee, barkan2023six,Barkan_2023_ICCV} or NLP~\cite{sikdar2021integrated,enguehard2023sequential}, recommender data is characterized by extreme sparsity and binary-valued interactions, where the absence of an interaction may be ambiguous. Standard PI methods, which integrate gradients from an all-zero baseline, fail in this setting due to weak or misleading attribution signals. Crucially, modern recommenders leverage both observed and unobserved interactions as informative signals. \method addresses this by stochastically sampling plausible user baselines from the empirical data distribution and selecting the explanation that maximizes fidelity. This adaptation enables more stable and faithful explanations tailored to the structure of recommender systems.


To evaluate \methodname, we conducted extensive fidelity evaluation spanning three model architectures (MF, VAE, NCF), multiple benchmark datasets (ML1M, Yahoo! Music, Pinterest), and a suite of counterfactual fidelity metrics~\citep{lxr,gurevitchlxr,mikhail_metrics}. Our results establish \method as the new state-of-the-art benchmark in recommender systems explainability, with ablation studies confirming the distinct contributions of both path-integration and our stochastic baseline sampling strategy.

\paragraph{Contributions.}
\begin{itemize}[leftmargin=*, itemsep=1pt, parsep=0pt, topsep=0pt]
    \item Introduce \methodname, the first adaptation of path-integration methods to recommender systems.
    \item Develop a novel stochastic baseline sampling strategy tailored to sparse, binary recommendation data.
    \item Conduct comprehensive fidelity-focused evaluation   across multiple architectures and datasets.
    \item Establish \methodname~as the new state-of-the-art for fidelity-aware recommendation explanations.
\end{itemize}
As fidelity remains an underexplored yet critical dimension in explainable recommendation~\cite{mikhail_metrics,Mohammadi2025beyond,koenigstein2025without}, we expect this work to lay essential groundwork for future research on trustworthy, model-faithful explanations.

\section{Related Work}
\label{sec:related}

The rapid growth of recommender systems has driven increasing interest in Explainable AI (XAI) methods to ensure transparency, build trust, and enhance user engagement~\cite{tintarev2022beyond,zhang2020explainable}. While a broad range of explanations for recommenders exist, standardized benchmarks for explanation fidelity remain significantly underexplored~\cite{mikhail_metrics,Mohammadi2025beyond,koenigstein2025without}.

\subsection{Explanation Methods for Recommenders}
Early works typically proposed model-specific explanations, such as those for matrix factorization~\citep{abdollahi2016explainable,abdollahi2017using} or inherently interpretable recommender architectures~\citep{barkan2020explainable,barkan2023modeling,melchiorre2022protomf,ai2vpp23,sugahara2024hierarchical}. Aspect-based methods attribute recommendations to human-interpretable item features (e.g., price or color)~\citep{vig2009tagsplanations,zhang2014explicit,wang2018explainable,li2021personalized}. However, these methods rely heavily on structured feature availability and are difficult to generalize to implicit or sparse data scenarios.

Model-agnostic explanation methods offer broader applicability by explaining arbitrary recommenders independently of their internal mechanisms. Prominent examples include LIME-RS~\citep{nobrega2019towards}, influence-based methods such as FIA and ACCENT~\citep{cheng2019incorporating,tran2021counterfactual}, and Shapley-value-based methods (SHAP4Rec and DeepSHAP)~\citep{shap4rec,deep_shap}. While model-agnostic methods enhance generality, their fidelity remains less scrutinized and insufficiently benchmarked.

\subsection{Fidelity Evaluation in Recommender Systems}
\label{sec:fidelity_in_recsys}
Explanation fidelity, the degree to which an explanation reflects the true reasoning of a recommender model, is essential for transparency and accountability, yet remains underexplored in the recommendation domain. Unlike computer vision or NLP, where fidelity evaluation is well-established~\citep{samek2016evaluating,agarwal2022openxai}, recommender systems have historically focused on user-centric goals like persuasiveness~\cite{tintarev2015explaining,zhang2020explainable} or satisfaction, often overlooking whether explanations faithfully reflect model logic~\citep{koenigstein2025without}.

Recent works have introduced counterfactual frameworks that evaluate fidelity by perturbing user histories and observing changes in recommendation outcomes~\citep{lxr,gurevitchlxr}. However, early approaches conflate supportive and contradictory features, apply coarse fixed-percentage masking, and lack control over explanation conciseness. The refined metrics proposed by~\citet{mikhail_metrics} address these limitations by evaluating fixed-length explanations, separating feature roles, and enabling consistent, interpretable comparisons across users.

Our work builds directly on this line of research, offering the first extensive empirical evaluation that spans both the original~\citep{lxr} and refined~\citep{mikhail_metrics} fidelity metrics. By benchmarking a wide range of explanation methods across multiple datasets and recommender models, we establish SPINRec as a new state-of-the-art for fidelity-aware explanations in recommender systems.

\subsection{Path-Integration (PI) for Explainability}

Path-integration (PI) techniques~\citep{sundararajan2017axiomatic} are widely adopted in computer vision and NLP~\citep{kapishnikov2021guided,xu2020attribution,sanyal2021discretized,enguehard2023sequential} to overcome limitations of vanilla gradients such as saturation and instability. By integrating gradients along a path from a baseline to the input, PI yields more robust and interpretable explanations.

However, directly applying existing PI methods to recommender systems is suboptimal. User representations in this domain are high-dimensional, sparse, and binary, where the absence of interaction conveys ambiguous information. Naïve baselines, such as all-zero “cold user” vectors, fail to reflect realistic user behavior and may produce weak or misleading gradient signals. This issue parallels challenges in computer vision, where black-image baselines distort attribution in dark regions~\citep{haug2021baselines}.


\method introduces a stochastic baseline sampling strategy explicitly tailored to these challenges. Instead of a single unrealistic baseline, it samples multiple plausible user histories from the empirical data distribution, capturing both presence and absence information. 

\section{The SPINRec Algorithm}
\label{sec:model}

We introduce \methodname~(Stochastic Path Integration for Neural Recommender Explanations), the first adaptation of path-integration methods to explain recommender systems. 

\subsection{ Setup and Notation}
\label{sec:model_setup}

Let \(\mathcal{U}\) and \(\mathcal{V}\) denote the sets of users and items, respectively. Each user \(u \in \mathcal{U}\) is represented by a binary feature vector \(\mathbf{x}_u \in \{0,1\}^{|\mathcal{V}|}\), where each feature \(\mathbf{x}_u[i] = 1\) indicates that user \(u\) interacted with item \(i\), and 0 otherwise.

We consider a recommender model \(f: \{0,1\}^{|\mathcal{V}|} \to [0,1]^{|\mathcal{V}|}\), parameterized by \(\theta\), which outputs predicted affinity scores over items given \(\mathbf{x}_u\). The predicted affinity for item \(y \in \mathcal{V}\) is denoted \(f^y(\mathbf{x}_u)\).

An explanation algorithm assigns a relevance score to each feature via an explanation map \(\mathbf{m} \in [0,1]^{|\mathcal{V}|}\), where \(\mathbf{m}[i]\) quantifies the contribution of feature \(i\) (i.e., $\x[i]$) to the prediction \(f^y(\mathbf{x}_u)\). High \(\mathbf{m}[i]\) values indicate stronger influence on the recommendation.

\subsection{Path-Integration for Recommenders}
\label{sec:path_int_methods}

Given a user data vector $\x$ and a recommended item $y$, \method attributes the predicted affinity score $\f^y(\x)$ to individual features in the user data vector by integrating gradients along a path from a baseline vector $\z$ to $\x$. We define a straight-line path $\r(t) = t \cdot \x + (1 - t) \cdot \z$ for $t \in [0,1]$, interpolating between the baseline and actual user representation.

The attribution is defined as the difference in predicted scores between the affinity of item to a user with personal data $\x$ and a baseline $\z$. The difference can be decomposed via the chain rule:
{\small
\begin{equation}
\label{eq:decomp}
\begin{split}
\f^y(\x) - \f^y(\z) 
&= \int_0^1 \frac{d}{dt} \f^y(\r(t)) \, dt 
= \int_0^1 \r'(t) \cdot \nabla \f^y(\r(t)) \, dt \\
&= \sum_{i=1}^{|\mathcal{V}|} \int_0^1 \frac{dr_i}{dt} \cdot \frac{\partial \f^y(\r(t))}{\partial r_i} \, dt,
\end{split}
\end{equation}
}
where $r_i(t)$ is the $i$-th coordinate of $\r(t)$.
Hence, an explanation map $\m$ can be calculated by attributing the difference between the predicted scores at $\x$ and $\z$ according to:
{\small
\begin{equation}
\label{eq:expl_map}
    \m = \int_0^1 \frac{\partial \f^y(\r(t))}{\partial \r(t)} \circ \frac{d\r(t)}{dt} \, dt,
\end{equation}
}
where $\circ$ denotes element-wise multiplication. 

While PI is well-studied in continuous domains, applying it effectively in sparse, binary recommender data requires careful baseline design.

\subsection{Challenges in Baseline Selection}
\label{sec:baseline_selection}
The choice of the baseline vector $\z$ significantly impacts the effectiveness of path-integration (PI) methods~\citep{haug2021baselines,erion2021improving}. While baseline sampling is common in other domains e.g., computer vision~\cite{erion2021improving}, recommender systems pose unique challenges due to the sparse and binary nature of user data:\begin{itemize}[noitemsep,leftmargin=*]
\item \textbf{Implicit Binary Signals:} Binary inputs limit the variability and restrict the range of values the model expects.
feedback restricts the applicability of continuous-valued baseline methods~\citep{sturmfels2020visualizing,haug2021baselines}.
\item \textbf{Data Sparsity:} Most user–item interactions are zeros, meaning that no meaningful path exists when integrating from an all-zero baseline.
\item \textbf{Diverse User Behaviors:} A single baseline may not capture the variability across user preferences and interactions.
\end{itemize}

Crucially, a naïve baseline (i.e., “cold user”) produces suboptimal gradient signals, since unobserved items remain zero during interpolation and thus contribute no gradients. However, modern recommenders leverage both observed and unobserved interactions as informative signals. This insight motivates our use of non-zero, data-driven baselines: by sampling plausible user profiles, \method captures both presence and absence effects, yielding significantly more faithful explanations, as confirmed by our ablation study.

\subsection{Stochastic Baseline Sampling}
\label{sec:stochastic_baseline}

\method introduces a stochastic sampling strategy tailored to recommender systems’ sparse and binary data. Instead of relying on a single baseline, it samples a set of $\kappa$ plausible baselines $\mathcal{B} = \{\z_1, \dots, \z_\kappa\}$ from the distribution of user histories, capturing the diversity and heterogeneity of user behavior.
For each $\z_i \in \mathcal{B}$, we compute an explanation map $\m_i$ using Eq.~\ref{eq:expl_map}. 
Then, the final explanation $\m^\ast$ is chosen to maximize a fidelity metric $s(\cdot)$ as follows:
{\small \begin{equation}
\m^\ast = \arg\max_{\m \in \mathcal{M} } s(\m), \label{eq:selection}
\end{equation}}
where $\mathcal{M} = \{\m_1, \dots, \m_\kappa\}$.
We note that beyond the maps in $\mathcal{M}$, one can further consider the mean map $\overline{\m} = \frac{1}{\kappa} \sum_{i=1}^{\kappa} \m_i$; by doing so, \method generalizes Expected Gradients~\citep{erion2021improving}.

Algorithm~\ref{alg:spinrec} summarizes the \method process, from baseline sampling to final explanation map selection.

\begin{algorithm}[H]
\small
\caption{\method: Stochastic Path-Integration}
\label{alg:spinrec}
\begin{algorithmic}[1]
\setlength{\itemsep}{0pt}
\STATE \textbf{Input:} User data $\x$, recommender $\f$, target item $y$, number of baselines to sample $\kappa$, metric $s$
\STATE \textbf{Output:} Explanation map $\mstar$
\STATE $\mathcal{M} \gets \{\}$; Sample $\kappa$ baselines from $\mathcal{U}$ to form the baselines set $\mathcal{B}$
\FOR{$\z \in \mathcal{B}$}
    \STATE Compute path \(\r(t)\) from $\z$ to $\x$
    \STATE Compute $\m$ via Eq.~\ref{eq:expl_map}; $\mathcal{M} \gets \mathcal{M} \cup \{\m\}$
\ENDFOR
\STATE \textbf{return} $\mstar \gets \underset{\m \in \mathcal{M} }{\operatorname{argmax}} \,s(\m)$
\end{algorithmic}
\end{algorithm}

\subsection{Computational Complexity}
\label{sec:complexity}

For each of the \(\kappa\) sampled baselines, \method integrates over \(J\) gradient steps, followed by \(N\) perturbation-based evaluations to compute $s(\cdot)$. Accordingly, \methodname's computational cost is dominated by two components: gradient-based integration (Eq.~\ref{eq:expl_map}) and the counterfactual evaluation via $s(\cdot)$ (Eq.~\ref{eq:selection}). 

For a model with \(Q\) parameters and \(|\mathcal{V}|\) items, the overall cost is:
{\small
\[
O\big(\kappa Q (J + N|\mathcal{V}|)\big) \approx O(\kappa Q N |\mathcal{V}|),
\]
}
since typically \(J \ll N|\mathcal{V}|\).

Compared to SHAP~\citep{shap4rec} with exponential cost in \(|\mathcal{V}|\), or LIME~\citep{lime_original} with cubic sample complexity, \method scales \emph{linearly} with the number of user features and baseline samples. All steps are embarrassingly parallel and well-suited to GPU acceleration. We note that while LXR~\citep{lxr} offers faster inference via a trained explainer, it requires pretraining and still falls short of \methodname’s fidelity as we show in our evaluations.

\section{Counterfactual Fidelity Metrics}
\label{sec:metrics}

As discussed earlier, recent work has introduced counterfactual fidelity metrics tailored to recommender systems~\citep{lxr,mikhail_metrics}. We build on this foundation by being the first to systematically evaluate both the original AUC-based metrics from \citet{lxr} and the refined fixed-length variants proposed by \citet{mikhail_metrics}. To ensure a fair comparison, we adhere strictly to the evaluation protocols used in these works. 

\begin{figure*}[t]
\centering
\includegraphics[scale=0.48]{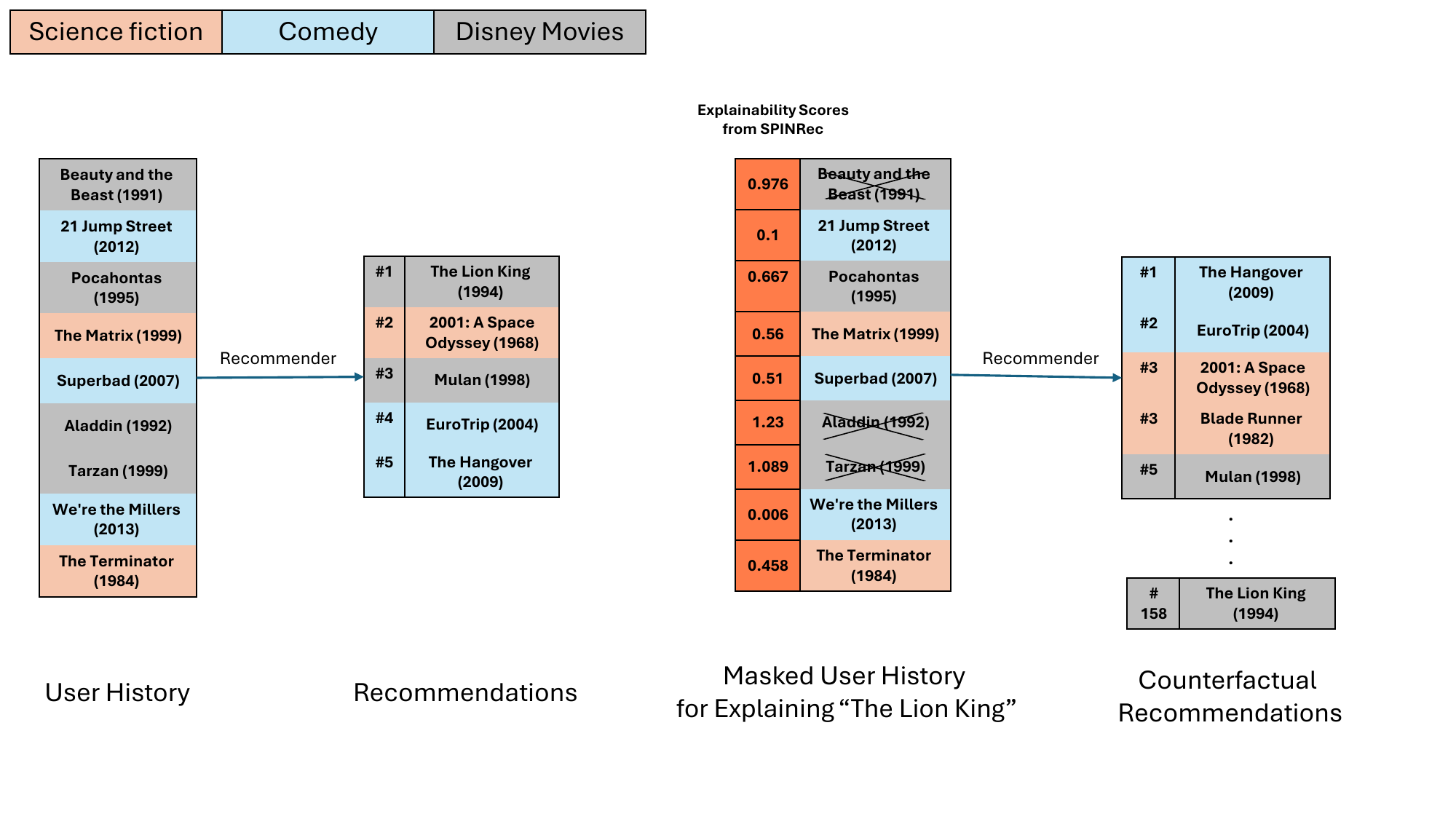}
\caption{\textbf{Illustration of Counterfactual Fidelity.} SPINRec identifies items in the user's history most responsible for recommending \emph{``The Lion King''}. When these items are masked, the recommendation's rank drastically drops, demonstrating explanation fidelity.}
\label{fig:example}
\end{figure*}

\paragraph{Illustrative Example.}
Figure~\ref{fig:example} illustrates the principal behind counterfactual fidelity evaluation. Given a user's interaction history, SPINRec identifies key features driving the recommendation of \emph{``The Lion King''}. Masking these features results in a substantial rank drop, empirically validating their explanatory power.

\paragraph{Formal Definitions.}
Let $\mathbf{x}_u$ denote user $u$'s historical interaction vector, and $K_e$ the number of top explanatory features. Define a binary mask $\mathbf{m}_{K_e}$ selecting the top $K_e$ features, and form two perturbed user vectors:

\textbf{Retained Explanations Vector:} $\xuretained = \xu \circ \mathbf{m}_{K_e}$

\textbf{Removed Explanations Vector:} $\xuremoved = \xu \circ (\mathbf{1} - \mathbf{m}_{K_e})$

The following metrics assess fidelity by measuring ranking or confidence changes for the target item $y$ under these counterfactual modifications:

\textbf{POS@$\boldsymbol{K_r, K_e}$:} Item $y$ drops out of top-$K_r$ recommendations when top-$K_e$ features are removed (lower is better):
\[
\text{POS@}K_r,K_e = \mathds{1}[\rankremoved \leq K_r].
\]

\textbf{DEL@$\boldsymbol{K_e}$:} Confidence drop after removing top-$K_e$ features (lower is better):
\[
\text{DEL@}K_e = \frac{f(\xuremoved)_y}{f(\xu)_y}.
\]

\textbf{INS@$\boldsymbol{K_e}$:} Confidence recovery from adding top-$K_e$ features (higher is better):
\[
\text{INS@}K_e = \frac{f(\xuretained)_y}{f(\xu)_y}.
\]

\textbf{CDCG@$\boldsymbol{K_e}$:} Rank degradation after removing explanatory features (lower is better):
\[
\text{CDCG@}K_e = \frac{1}{\log_2(1+\rankremoved)}.
\]
\paragraph{AUC Computation.}
To compute AUC variants, we follow the fixed-step perturbation strategy of \citet{lxr}, which averages the model’s scores as features in the user vector are progressively removed or added.

\section{Experimental Setup}
\label{sec:experimental-setup}

Our setup builds on the protocol of \citet{lxr}, extending it with a third dataset (Pinterest), additional fidelity metrics from \citet{mikhail_metrics}, and a broader set of explanation baselines. Due to space constraints, the Pinterest results are provided in our public repository. Hyperparameters were tuned via grid search on a held-out validation set, with final values also available in the repository. All experiments were conducted on NVIDIA V100 GPUs using PyTorch~1.13 and CUDA~11.7.

\subsection{Recommendation Models} 
\label{sec:eval-recommenders} 

We evaluate \methodname\ across three standard recommendation models:

\textbf{Matrix Factorization (MF)}~\citep{koren2009matrix}: Despite its simplicity, MF remains competitive with modern recommenders~\citep{rendle2022revisiting}. We use a dynamic variant that derives user embeddings directly from interaction vectors.

\textbf{Variational Autoencoder (VAE)}~\citep{liang2018variational,shenbin2020recvae}: A generative latent variable model that reconstructs user-item vectors from compressed representations.

\textbf{Neural Collaborative Filtering (NCF)}~\citep{he2017neural}: A hybrid architecture combining matrix factorization and multi-layer perceptrons to model nonlinear user-item interactions.

\subsection{Datasets}
\label{sec:eval-datasets}

Experiments were conducted on three datasets:
\textbf{ML1M}~\citep{harper2015movielens}, 
\textbf{Yahoo! Music}~\citep{dror2012yahoo}, and 
\textbf{Pinterest}~\citep{he2017neural}. All datasets were binarized to implicit feedback, with an 80/20 user-based train-test split. An additional 10\% of users were withheld from training for hyperparameter tuning. All results are reported on the test set, where explanations target the top recommendation per user.  

\subsection{Baselines and Methods}
\label{sec:eval-methods}

We compare \method against a broad set of post-hoc, model-agnostic explanation baselines, spanning heuristic, perturbation-based, and learning-based methods:

\textbf{Cosine Similarity:} A non-counterfactual heuristic that ranks user-history items by cosine similarity to the recommended item~\citep{singh2020movie}.

\textbf{SHAP4Rec}~\citep{shap4rec}: A perturbation-based method grounded in Shapley values~\citep{shap_original}, adapted for recommendation via Jaccard-based clustering and \(K=10\) k-means sampling, as in~\citep{lxr}.

\textbf{DeepSHAP}~\citep{deep_shap}: A fast SHAP approximation using DeepLIFT-style gradient propagation~\citep{shrikumar2017learning}.

\textbf{LIME-RS}~\citep{nobrega2019towards}: A LIME adaptation for recommender systems, fitting a local linear surrogate model around a perturbed user profile.

\textbf{LIRE}~\citep{brunot2022preference}: A robust LIME variant using importance sampling to improve faithfulness in sparse recommendation domains.

\textbf{FIA}~\citep{cheng2019incorporating}: An approach utilizing influence functions to estimate the effect of each user feature. 

\textbf{ACCENT}~\citep{tran2021counterfactual}: A fidelity-aware explainer based on influence functions~\citep{koh2017understanding}, extending FIA to capture second-order model effects.

\textbf{LXR}~\citep{lxr}: A state-of-the-art fidelity-aware method that learns an auxiliary explainer network to optimize counterfactual metrics under perturbation.

\textbf{PI (Ablated \method):} A vanilla path-integration baseline that omits the stochastic baseline sampling of \method. This model serves to isolate the contribution of sampling, showing that while PI alone achieves strong fidelity, it remains suboptimal. The full \method significantly outperforms ABLT across all settings, demonstrating the importance of adapting PI to sparse recommender data.

\textbf{\method (Ours):} The proposed method combines path integration with fidelity-optimized stochastic baseline sampling to generate high-precision attribution maps for recommendation outcomes.


\begin{table*}[]
\centering
\begin{minipage}[t]{0.48\textwidth}
\centering
\captionof{table}{ML1M Dataset}
\label{tab:ML1M}
\resizebox{\linewidth}{!}{
\begin{tabular}{|c||l||c|c|c|c|c|c|}
\hline
\multirow{1}{*}{Rec} & Method & POS@5 $\downarrow$ & POS@10 $\downarrow$ & POS@20 $\downarrow$ & DEL $\downarrow$ & INS $\uparrow$ & CDCG $\downarrow$ \\
\hline
\multirow{10}{*}{MF} 
& Cosine     & 0.646 & 0.703 & 0.744 & 0.776 & 0.911 & 0.589 \\
& SHAP       & 0.812 & 0.857 & 0.883 & 0.851 & 0.858 & 0.734 \\
& DeepSHAP   & 0.431 & 0.499 & 0.547 & \underline{0.564} & \underline{0.938} & 0.422 \\
& LIME       & 0.644 & 0.725 & 0.778 & 0.735 & 0.928 & 0.576 \\
& LIRE       & 0.588 & 0.651 & 0.694 & 0.678 & 0.926 & 0.512 \\
& FIA        & 0.432 & 0.497 & {0.543} & {0.570} & \underline{0.938} & {0.422} \\
& ACCENT     & 0.707 & 0.760 & 0.797 & 0.729 & 0.910 & 0.652 \\
& LXR        & 0.457 & 0.521 & 0.571 & 0.593 & 0.936 & 0.442 \\
& PI       & \underline{0.418} & \underline{0.483} & \underline{0.529} & \textbf{0.555} & \textbf{0.939} & \underline{0.411} \\
& SPINRec    & \textbf{0.410} & \textbf{0.478} & \textbf{0.527} & \textbf{0.555} & \textbf{0.939} & \textbf{0.405} \\
\hline
\multirow{10}{*}{VAE}
& Cosine     & 0.412 & 0.501 & 0.595 & 0.007 & 0.020 & 0.435 \\
& SHAP       & 0.602 & 0.689 & 0.766 & 0.011 & 0.011 & 0.572 \\
& DeepSHAP   & 0.340 & 0.410 & 0.490 & 0.007 & 0.025 & 0.396 \\
& LIME       & 0.502 & 0.604 & 0.698 & 0.008 & 0.016 & 0.502 \\
& LIRE       & 0.345 & 0.437 & 0.536 & 0.007 & 0.021 & 0.392 \\
& FIA        & \underline{0.234} & \underline{0.312} & \underline{0.411} & \textbf{0.005} & \underline{0.029} & \underline{0.320} \\
& ACCENT     & 0.483 & 0.565 & 0.649 & 0.007 & 0.017 & 0.505 \\
& LXR        & 0.348 & 0.430 & 0.518 & \underline{0.006} & 0.022 & 0.394 \\
& PI       & 0.236 & 0.319 & 0.416 & \textbf{0.005} & \underline{0.029} & 0.322 \\
& SPINRec    & \textbf{0.189} & \textbf{0.252} & \textbf{0.335} & \textbf{0.005} & \textbf{0.031} & \textbf{0.293} \\
\hline
\multirow{10}{*}{NCF}
& Cosine     & 0.263 & 0.315 & 0.360 & 0.485 & 0.769 & 0.312 \\
& SHAP       & 0.501 & 0.559 & 0.602 & 0.620 & 0.667 & 0.504 \\
& DeepSHAP   & \underline{0.210} & \underline{0.247} & \underline{0.282} & \underline{0.387} & \underline{0.805} & \underline{0.272} \\
& LIME       & 0.291 & 0.345 & 0.391 & 0.484 & 0.774 & 0.334 \\
& LIRE       & 0.301 & 0.350 & 0.397 & 0.474 & 0.776 & 0.338 \\
& FIA        & 0.215 & 0.256 & 0.293 & 0.591 & \underline{0.805} & 0.276 \\
& ACCENT     & 0.306 & 0.348 & 0.387 & 0.462 & 0.774 & 0.347 \\
& LXR        & 0.249 & 0.301 & 0.346 & 0.451 & 0.790 & 0.303 \\
& PI       & 0.211 & 0.248 & 0.283 & 0.389 & 0.807 & 0.273 \\
& SPINRec    & \textbf{0.185} & \textbf{0.223} & \textbf{0.261} & \textbf{0.382} & \textbf{0.810} & \textbf{0.258} \\
\hline
\end{tabular}
}

\end{minipage}
\hfill
\begin{minipage}[t]{0.48\textwidth}
\centering
\captionof{table}{Yahoo Dataset}
\label{tab:Yahoo}
\resizebox{\linewidth}{!}{
\begin{tabular}{|c||l||c|c|c|c|c|c|}
\hline
\multirow{1}{*}{Rec} & Method & POS@5 $\downarrow$ & POS@10 $\downarrow$ & POS@20 $\downarrow$ & DEL $\downarrow$ & INS $\uparrow$ & CDCG $\downarrow$ \\
\hline
\multirow{12}{*}{MF}
& Cosine     & 0.331 & 0.436 & 0.504 & 0.695 & 0.868 & 0.382 \\
& SHAP       & 0.530 & 0.637 & 0.697 & 0.765 & 0.821 & 0.533 \\
& DeepSHAP   & 0.258 & 0.348 & 0.401 & \underline{0.592} & \underline{0.882} & 0.324 \\
& LIME       & 0.360 & 0.463 & 0.530 & 0.681 & 0.871 & 0.402 \\
& LIRE       & 0.424 & 0.525 & 0.581 & 0.701 & 0.858 & 0.448 \\
& FIA        & 0.263 & 0.352 & 0.408 & 0.601 & \underline{0.882} & 0.328 \\
& ACCENT     & 0.364 & 0.452 & 0.505 & 0.646 & 0.874 & 0.411 \\
& LXR        & 0.282 & 0.371 & 0.428 & 0.620 & 0.879 & 0.344 \\
& PI       & \underline{0.256} & \underline{0.344} & \underline{0.398} & \textbf{0.591} & \textbf{0.883} & \underline{0.323} \\
& SPINRec    & \textbf{0.246} & \textbf{0.337} & \textbf{0.393} & \textbf{0.591} & \textbf{0.883} & \textbf{0.318} \\
\hline
\multirow{12}{*}{VAE}
& Cosine     & 0.402 & 0.503 & 0.605 & 0.014 & 0.042 & 0.432 \\
& SHAP       & 0.605 & 0.690 & 0.766 & 0.021 & 0.031 & 0.576 \\
& DeepSHAP   & 0.362 & 0.454 & 0.558 & 0.014 & 0.043 & 0.410 \\
& LIME       & 0.576 & 0.664 & 0.744 & 0.021 & 0.032 & 0.554 \\
& LIRE       & 0.453 & 0.559 & 0.665 & 0.017 & 0.038 & 0.463 \\
& FIA        & \underline{0.315} & \underline{0.420} & \underline{0.542} & \underline{0.013} & \underline{0.048} & \underline{0.377} \\
& ACCENT     & 0.541 & 0.622 & 0.704 & 0.019 & 0.032 & 0.553 \\
& LXR        & 0.393 & 0.488 & 0.589 & 0.014 & 0.041 & 0.429 \\
& PI       & 0.320 & 0.424 & 0.545 & \underline{0.013} & \underline{0.048} & 0.380 \\
& SPINRec    & \textbf{0.280} & \textbf{0.376} & \textbf{0.489} & \textbf{0.012} & \textbf{0.049} & \textbf{0.358} \\
\hline
\multirow{12}{*}{NCF}
& Cosine     & 0.456 & 0.512 & 0.579 & 0.619 & 0.723 & 0.466 \\
& SHAP       & 0.561 & 0.603 & 0.657 & 0.657 & 0.683 & 0.559 \\
& DeepSHAP   & 0.253 & 0.285 & 0.339 & \underline{0.526} & \underline{0.768} & 0.315 \\
& LIME       & 0.290 & 0.328 & 0.386 & 0.549 & 0.762 & 0.347 \\
& LIRE       & 0.417 & 0.460 & 0.517 & 0.590 & 0.736 & 0.443 \\
& FIA        & 0.254 & 0.288 & 0.342 & 0.527 & \underline{0.768} & 0.317 \\
& ACCENT     & 0.312 & 0.346 & 0.399 & 0.547 & 0.758 & 0.366 \\
& LXR        & 0.267 & 0.300 & 0.353 & 0.536 & 0.764 & 0.327 \\
& PI       & \underline{0.250} & \underline{0.284} & \underline{0.338} & 0.527 & \underline{0.768} & \underline{0.313} \\
& SPINRec    & \textbf{0.238} & \textbf{0.273} & \textbf{0.328} & \textbf{0.525} & \textbf{0.769} & \textbf{0.305} \\
\hline
\end{tabular}
}

\end{minipage}
\label{tab:ml1m_yahoo}
\end{table*}

\section{Counterfactual Evaluation Results}
\label{sec:results}

We evaluate \method across three recommender architectures (MF, VAE, NCF) and three benchmark datasets (ML1M, Yahoo! Music, Pinterest), using both AUC-style~\citep{lxr} and fixed-length~\citep{mikhail_metrics} counterfactual fidelity metrics. Results for Pinterest are reported in our public repository.

\paragraph{AUC-Based Metrics.}
Tables~\ref{tab:ML1M}--\ref{tab:Yahoo} report Area-Under-Curve (AUC) scores, summarizing fidelity degradation under stepwise perturbations. Across all models and datasets, \method achieves the best results, significantly surpassing strong baselines such as LXR and FIA ($p \leq 0.01$, paired t-test). 

\paragraph{Fixed-Length Fidelity Metrics.}
Tables~\ref{tab:fidelity_metrics_all}--\ref{tab:fidelity_metrics_yahoo} present fidelity at fixed explanation lengths \(K_e \in \{2,3,4\}\) and ranking cutoffs \(K_r \in \{5,10,20\}\), simulating realistic user-facing scenarios where concise, high-impact explanations are crucial. As \(K_r\) increases and \(K_e\) decreases, the counterfactual test becomes more challenging: fewer explanatory items must shift the recommended item beyond a stricter cutoff. This difficulty is reflected in tighter performance margins, especially at \(K_e{=}2\) and \(3\), where multiple methods sometimes tie. We omit \(K_e{=}1\) here due to its instability and limited discriminative power, but include results for \(K_e{=}1\) and \(5\) in our public repository, where trends remain consistent. \method again outperforms all baselines across all configurations, confirming its robustness across fidelity granularities.


\subsection{Ablation Study: Plain vs.\ Stochastic Path Integration}
To isolate the contribution of stochastic baseline sampling, we compare \method to its ablated variant (\textbf{PI}), which employs plain path integration without sampling. While PI performs competitively and often ranks near the top, \method consistently achieves superior fidelity across most metrics and datasets. This improvement reflects a key insight: modern recommenders rely not only on observed interactions but also on their absence as informative signals. By sampling diverse, non-zero baselines, \method allows missing items to contribute meaningful gradients. Gains over PI are most pronounced in advanced models (VAE, NCF), where unobserved interactions play a larger role.

\begin{figure}[]
\centering
\includegraphics[width=0.43\textwidth, keepaspectratio]{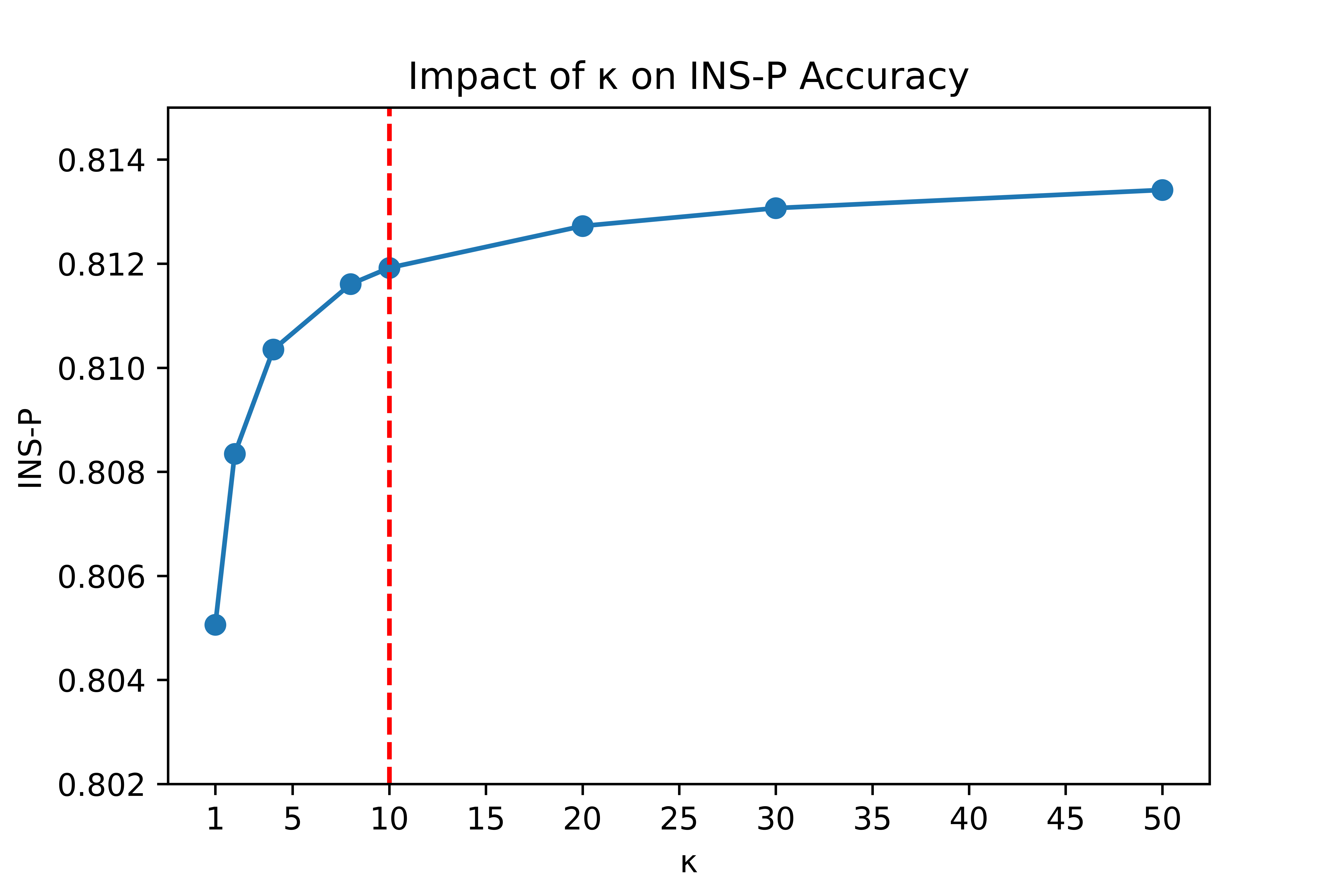}
\caption{Fidelity (INS) vs. number of baseline samples ($\kappa$) for NCF on ML1M. Gains plateau after $\kappa{=}10$.}
\label{fig:10_ml1m_vae_ins}
\end{figure}

\paragraph{Impact of Sampling Count \(\kappa\).}
\label{sec:diminishing_returns}
Figure~\ref{fig:10_ml1m_vae_ins} shows the effect of varying the number of sampled baselines \(\kappa\) on explanation fidelity. Performance plateaus at around \(\kappa{=}10\), indicating a strong balance between fidelity and computational efficiency.

\paragraph{Summary and Insights.}
Across all metrics, datasets, and recommender architectures, \method consistently outperforms existing explanation methods, establishing a new fidelity benchmark for recommendation. Path integration (PI) itself proves to be an inherently strong approach for fidelity, even without stochastic sampling. Incorporating stochastic baseline sampling further enhances fidelity by leveraging both observed and unobserved interactions, an effect particularly pronounced in more expressive models such as VAE and NCF.

\section{Conclusion}
\label{sec:conclusion}

We introduced \methodname, the first model-agnostic explanation method to apply path integration (PI) to recommender systems. By combining PI with a stochastic baseline sampling strategy tailored to sparse, binary user–item data, \methodname produces stable, high-fidelity explanations that more faithfully capture model reasoning.

A comprehensive evaluation across multiple models, datasets, and fidelity metrics demonstrates that \methodname consistently outperforms existing approaches, establishing a new benchmark for fidelity-aware explainability in recommendation.

We hope this work encourages further exploration of fidelity-focused explainability in recommendation.

\begin{table*}[]
\centering
\small
\caption{Fidelity Metrics at Different \(K_r, K_e\) Values for MF, VAE, and NCF (ML1M)}
\label{tab:fidelity_metrics_all}
\resizebox{\linewidth}{!}{
\begin{tabular}{|l||ccc|ccc|ccc|ccc|ccc|ccc|}
\hline
\multirow{3}{*}{Method} 
& \multicolumn{3}{c|}{POS@5,\(K_e\) $\downarrow$} 
& \multicolumn{3}{c|}{POS@10,\(K_e\) $\downarrow$} 
& \multicolumn{3}{c|}{POS@20,\(K_e\) $\downarrow$} 
& \multicolumn{3}{c|}{DEL@\(K_e\) $\downarrow$} 
& \multicolumn{3}{c|}{INS@\(K_e\) $\uparrow$} 
& \multicolumn{3}{c|}{CDCG@\(K_e\) $\downarrow$} \\
& \(K_e{=}2\) & \(K_e{=}3\) & \(K_e{=}4\) 
& \(K_e{=}2\) & \(K_e{=}3\) & \(K_e{=}4\) 
& \(K_e{=}2\) & \(K_e{=}3\) & \(K_e{=}4\) 
& \(K_e{=}2\) & \(K_e{=}3\) & \(K_e{=}4\) 
& \(K_e{=}2\) & \(K_e{=}3\) & \(K_e{=}4\) 
& \(K_e{=}2\) & \(K_e{=}3\) & \(K_e{=}4\) \\
\hline
\multicolumn{19}{|c|}{\textbf{MF}} \\
\hline
Cosine   & 0.987 & 0.964 & 0.945 & 0.996 & 0.988 & 0.974 & \underline{0.998} & 0.993 & 0.988 & 0.983 & 0.974 & 0.965 & 0.750 & 0.785 & 0.812 & 0.915 & 0.883 & 0.861 \\
SHAP     & 1.000 & 0.999 & 0.994 & 1.000 & 0.999 & 0.998 & 1.000 & 1.000 & 1.000 & 0.997 & 0.995 & 0.992 & 0.648 & 0.662 & 0.676 & 0.969 & 0.956 & 0.945 \\
DeepSHAP & \underline{0.980} & \underline{0.945} & 0.921 & \textbf{0.993} & 0.974 & 0.957 & \textbf{0.997} & \underline{0.991} & \underline{0.978} & \underline{0.975} & \underline{0.960} & \underline{0.945} & 0.795 & 0.842 & 0.877 & 0.886 & 0.836 & 0.791 \\
LIME     & 0.987 & 0.964 & 0.948 & \underline{0.994} & 0.988 & 0.976 & \textbf{0.997} & 0.993 & 0.988 & 0.980 & 0.968 & 0.958 & 0.751 & 0.788 & 0.818 & 0.916 & 0.884 & 0.850 \\
LIRE     & 0.993 & 0.974 & 0.964 & 0.999 & 0.993 & 0.985 & 1.000 & 0.997 & 0.990 & 0.985 & 0.977 & 0.969 & 0.764 & 0.804 & 0.836 & 0.926 & 0.894 & 0.871 \\
FIA      & \underline{0.980} & \underline{0.945} & \underline{0.916} & \textbf{0.993} & \underline{0.974} & \underline{0.953} & \textbf{0.997} & \textbf{0.988} & \textbf{0.974} & \textbf{0.974} & \textbf{0.959} & \textbf{0.944} & 0.796 & 0.842 & 0.876 & \underline{0.884} & \underline{0.831} & \underline{0.788} \\
ACCENT   & 0.994 & 0.984 & 0.977 & 0.998 & 0.993 & 0.987 & \underline{0.998} & 0.995 & 0.991 & 0.983 & 0.973 & 0.962 & 0.750 & 0.786 & 0.815 & 0.953 & 0.933 & 0.910 \\
LXR      & 0.988 & 0.958 & 0.933 & 0.997 & 0.980 & 0.962 & 1.000 & 0.993 & \underline{0.978} & 0.979 & 0.966 & 0.951 & 0.779 & 0.828 & 0.865 & 0.904 & 0.862 & 0.820 \\
PI     & \underline{0.980} & \underline{0.945} & \underline{0.916} & \textbf{0.993} & \underline{0.974} & \underline{0.953} & \textbf{0.997 }& \textbf{0.988 }& \textbf{0.974} & \textbf{0.974 }& \textbf{0.959} & \textbf{0.944 }& \underline{0.798} & \underline{0.845} & \underline{0.879} & \underline{0.884} & 0.832 & 0.789 \\
SPINRec  & \textbf{0.975} & \textbf{0.940} & \textbf{0.911} & \textbf{0.993} & \textbf{0.972 }& \textbf{0.952} & \textbf{0.997 }& \textbf{0.988} & \textbf{0.974} & \textbf{0.974} & \textbf{0.959 }& \textbf{0.944} & \textbf{0.799} & \textbf{0.846} & \textbf{0.880} & \textbf{0.859} & \textbf{0.806} & \textbf{0.762 }\\
\hline
\multicolumn{19}{|c|}{\textbf{VAE}} \\
\hline
Cosine   & 0.976 & 0.942 & 0.906 & 0.993 & 0.982 & 0.961 & 0.998 & 0.995 & 0.986 & 0.895 & 0.856 & 0.821 & 2.807 & 3.125 & 3.286 & 0.861 & 0.825 & 0.782 \\
SHAP     & 0.994 & 0.984 & 0.978 & 0.999 & 0.997 & 0.989 & 0.999 & 0.998 & 0.997 & 0.986 & 0.983 & 0.973 & 0.617 & 0.656 & 0.696 & 0.941 & 0.922 & 0.903 \\
DeepSHAP & 0.952 & 0.903 & 0.840 & 0.986 & 0.964 & 0.929 & \underline{0.997} & 0.989 & 0.970 & 0.876 & 0.832 & 0.795 & 2.162 & 2.536 & 2.770 & 0.828 & 0.772 & 0.723 \\
LIME     & 0.985 & 0.959 & 0.938 & 0.996 & 0.991 & 0.973 & 0.999 & 0.998 & 0.993 & 0.936 & 0.911 & 0.886 & 1.015 & 1.108 & 1.216 & 0.893 & 0.863 & 0.830 \\
LIRE     & 0.967 & 0.942 & 0.909 & 0.991 & 0.980 & 0.969 & 0.998 & 0.995 & 0.990 & 0.878 & 0.836 & 0.800 & 1.588 & 1.836 & 2.014 & 0.841 & 0.784 & 0.737 \\
FIA      & \underline{0.921} & \underline{0.844} & \underline{0.752} & \underline{0.978} & \underline{0.937} & \underline{0.891} & \textbf{0.996} & \underline{0.983} & \underline{0.960} & \underline{0.810} & \underline{0.746} & \underline{0.694} & 2.002 & 2.429 & 2.703 & 0.763 & \underline{0.680} & \underline{0.616} \\
ACCENT   & 0.988 & 0.968 & 0.948 & 0.998 & 0.983 & 0.974 & 0.998 & 0.995 & 0.986 & 0.901 & 0.864 & 0.831 & 0.998 & 1.139 & 1.246 & 0.945 & 0.920 & 0.895 \\
LXR      & 0.983 & 0.944 & 0.906 & 0.995 & 0.979 & 0.959 & 0.999 & 0.994 & 0.982 & 0.906 & 0.864 & 0.826 & \underline{2.823} & \underline{3.369} & \underline{3.873} & 0.877 & 0.835 & 0.782 \\
PI     & 0.922 & 0.848 & 0.770 & 0.981 & \underline{0.937} & 0.902 & \underline{0.997} & 0.984 & 0.963 & 0.814 & 0.752 & 0.701 & 1.987 & 2.400 & 2.676 & \underline{0.761} & 0.689 & 0.626 \\
SPINRec  & \textbf{0.903} & \textbf{0.791} & \textbf{0.687} & \textbf{0.972} & \textbf{0.923} & \textbf{0.857} & \textbf{0.996} & \textbf{0.978} & \textbf{0.945} & \textbf{0.808} & \textbf{0.741} & \textbf{0.685} & \textbf{3.626} & \textbf{4.197} & \textbf{4.524} & \textbf{0.703} & \textbf{0.616} & \textbf{0.550} \\
\hline
\multicolumn{19}{|c|}{\textbf{NCF}} \\
\hline
Cosine   & 0.906 & 0.835 & 0.762 & 0.964 & 0.917 & 0.845 & 0.983 & 0.945 & 0.908 & 0.947 & 0.922 & 0.897 & 0.572 & 0.637 & 0.696 & 0.811 & 0.740 & 0.674 \\
SHAP     & 0.979 & 0.955 & 0.930 & 0.998 & 0.997 & 0.983 & 1.000 & 0.993 & 0.983 & 0.985 & 0.977 & 0.968 & 0.480 & 0.507 & 0.533 & 0.910 & 0.879 & 0.845 \\
DeepSHAP & 0.908 & 0.820 & 0.728 & 0.986 & 0.964 & 0.929 & 0.997 & 0.938 & 0.887 & 0.936 & 0.905 & 0.875 & \underline{0.600} & \underline{0.673} & \underline{0.739} & 0.788 & 0.711 & 0.634 \\
LIME     & 0.939 & 0.884 & 0.823 & 0.971 & 0.934 & 0.887 & 0.984 & 0.959 & 0.923 & 0.948 & 0.924 & 0.901 & 0.560 & 0.615 & 0.666 & 0.836 & 0.772 & 0.717 \\
LIRE     & 0.935 & 0.879 & 0.825 & 0.974 & 0.932 & 0.895 & 0.988 & 0.966 & 0.937 & 0.947 & 0.924 & 0.902 & 0.557 & 0.613 & 0.663 & 0.824 & 0.765 & 0.712 \\
FIA      & \underline{0.887} & \underline{0.792} & \underline{0.699} & \underline{0.945} & \underline{0.887} & \underline{0.806} & \underline{0.971} & \underline{0.930} & \underline{0.882} & \textbf{0.929} & \underline{0.896} & \underline{0.865} & 0.586 & 0.655 & 0.715 & \underline{0.775} & \underline{0.685} & \underline{0.615} \\
ACCENT   & 0.957 & 0.921 & 0.871 & 0.981 & 0.955 & 0.916 & 0.991 & 0.970 & 0.954 & 0.944 & 0.917 & 0.892 & 0.556 & 0.612 & 0.664 & 0.892 & 0.840 & 0.791 \\
LXR      & 0.923 & 0.840 & 0.737 & 0.959 & 0.911 & 0.842 & 0.976 & 0.943 & 0.899 & 0.938 & 0.908 & 0.877 & 0.587 & 0.659 & 0.727 & 0.813 & 0.732 & 0.661 \\
PI     & 0.893 & 0.799 & 0.705 & 0.947 & 0.890 & 0.814 & 0.975 & 0.935 & 0.890 & \underline{0.932} & 0.901 & 0.870 & 0.585 & 0.654 & 0.716 & 0.782 & 0.694 & 0.621 \\
SPINRec  & \textbf{0.864} & \textbf{0.747} & \textbf{0.637} & \textbf{0.937} & \textbf{0.866} & \textbf{0.772} & \textbf{0.969} & \textbf{0.916} & \textbf{0.858} & \textbf{0.929} & \textbf{0.895} & \textbf{0.863} & \textbf{0.612} & \textbf{0.689} & \textbf{0.757} & \textbf{0.723} & \textbf{0.629} & \textbf{0.560} \\
\hline
\end{tabular}
}
\end{table*}
\begin{table*}[]
\centering
\small
\caption{Fidelity Metrics at Different \(K_r, K_e\) Values for MF, VAE, and NCF on Yahoo!}
\label{tab:fidelity_metrics_yahoo}

\resizebox{\linewidth}{!}{
\begin{tabular}{|l||ccc|ccc|ccc|ccc|ccc|ccc|}
\hline
\multirow{3}{*}{Method} 
& \multicolumn{3}{c|}{POS@5,\(K_e\) $\downarrow$} 
& \multicolumn{3}{c|}{POS@10,\(K_e\) $\downarrow$} 
& \multicolumn{3}{c|}{POS@20,\(K_e\) $\downarrow$} 
& \multicolumn{3}{c|}{DEL@\(K_e\) $\downarrow$} 
& \multicolumn{3}{c|}{INS@\(K_e\) $\uparrow$} 
& \multicolumn{3}{c|}{CDCG@\(K_e\) $\downarrow$} \\
& \(K_e{=}2\) & \(K_e{=}3\) & \(K_e{=}4\) 
& \(K_e{=}2\) & \(K_e{=}3\) & \(K_e{=}4\) 
& \(K_e{=}2\) & \(K_e{=}3\) & \(K_e{=}4\) 
& \(K_e{=}2\) & \(K_e{=}3\) & \(K_e{=}4\) 
& \(K_e{=}2\) & \(K_e{=}3\) & \(K_e{=}4\) 
& \(K_e{=}2\) & \(K_e{=}3\) & \(K_e{=}4\) \\
\hline
\multicolumn{19}{|c|}{\textbf{MF}} \\
\hline
Cosine     & 0.789 & 0.672 & 0.586 & 0.872 & 0.786 & 0.723 & 0.911 & 0.845 & 0.795 & 0.926 & 0.895 & 0.869 & 0.769 & 0.839 & 0.885 & 0.687 & 0.597 & 0.532 \\
SHAP       & 0.879 & 0.811 & 0.754 & 0.927 & 0.883 & 0.843 & 0.947 & 0.915 & 0.879 & 0.949 & 0.925 & 0.905 & 0.655 & 0.709 & 0.751 & 0.811 & 0.748 & 0.703 \\
DeepSHAP   & 0.745 & 0.619 & 0.501 & 0.839 & 0.749 & 0.662 & 0.889 & 0.817 & 0.744 & \underline{0.912} & 0.875 & \underline{0.843} & \underline{0.826} & \underline{0.897} & \underline{0.939} & 0.643 & 0.549 & 0.476 \\
LIME       & 0.803 & 0.704 & 0.622 & 0.882 & 0.804 & 0.747 & 0.912 & 0.854 & 0.810 & 0.923 & 0.891 & 0.864 & 0.772 & 0.840 & 0.886 & 0.714 & 0.630 & 0.567 \\
LIRE       & 0.862 & 0.776 & 0.711 & 0.918 & 0.864 & 0.818 & 0.939 & 0.903 & 0.868 & 0.942 & 0.916 & 0.892 & 0.746 & 0.814 & 0.864 & 0.762 & 0.694 & 0.635 \\
FIA        & \underline{0.743} & 0.610 & \underline{0.492} & \underline{0.837} & \underline{0.744} & \underline{0.652} & 0.887 & \underline{0.812} & \underline{0.738} & \textbf{0.911} & \textbf{0.874} & \textbf{0.841} & 0.825 & 0.895 & 0.936 & \underline{0.638} & \underline{0.545} & \underline{0.475} \\
ACCENT     & 0.833 & 0.753 & 0.676 & 0.888 & 0.826 & 0.775 & 0.918 & 0.867 & 0.824 & 0.920 & 0.888 & 0.860 & 0.788 & 0.857 & 0.902 & 0.778 & 0.704 & 0.641 \\
LXR        & 0.778 & 0.652 & 0.541 & 0.859 & 0.769 & 0.687 & 0.900 & 0.832 & 0.763 & 0.917 & \underline{0.881} & 0.849 & 0.813 & 0.886 & 0.930 & 0.678 & 0.581 & 0.506 \\
PI       & \underline{0.743} & \underline{0.609} & \underline{0.492} & \underline{0.837} & \underline{0.744} & \underline{0.652} & \textbf{0.879} & 0.813 & \underline{0.738} & \textbf{0.911} & \textbf{0.874} & \textbf{0.841} & 0.825 & 0.894 & 0.934 & \underline{0.638} & 0.546 & 0.476 \\
SPINRec    & \textbf{0.722} & \textbf{0.579} & \textbf{0.458} & \textbf{0.829} & \textbf{0.731} & \textbf{0.633} & \underline{0.884} & \textbf{0.806} & \textbf{0.729} & \textbf{0.911} & \textbf{0.874} & \textbf{0.841} & \textbf{0.829} & \textbf{0.899} & \textbf{0.940} & \textbf{0.608 }& \textbf{0.512} & \textbf{0.442} \\
\hline
\multicolumn{19}{|c|}{\textbf{VAE}} \\
\hline
Cosine     & 0.834 & 0.741 & 0.671 & 0.908 & 0.844 & 0.782 & 0.943 & 0.897 & 0.854 & 0.739 & 0.657 & 0.597 & \underline{2.026} & 2.093 & 2.103 & 0.716 & 0.637 & 0.577 \\
SHAP       & 0.900 & 0.848 & 0.804 & 0.938 & 0.895 & 0.857 & 0.957 & 0.925 & 0.894 & 0.886 & 0.836 & 0.793 & 0.711 & 0.785 & 0.849 & 0.825 & 0.770 & 0.725 \\
DeepSHAP   & 0.813 & 0.702 & 0.614 & 0.895 & 0.820 & 0.752 & 0.939 & 0.887 & 0.838 & 0.737 & 0.653 & 0.590 & 1.785 & 1.885 & 1.927 & 0.687 & 0.600 & 0.538 \\
LIME       & 0.897 & 0.836 & 0.791 & 0.934 & 0.897 & 0.854 & 0.991 & 0.926 & 0.896 & 0.885 & 0.831 & 0.788 & 0.921 & 1.015 & 1.077 & 0.809 & 0.754 & 0.710 \\
LIRE       & 0.875 & 0.803 & 0.738 & 0.930 & 0.881 & 0.832 & 0.956 & 0.922 & 0.886 & 0.805 & 0.731 & 0.672 & 1.319 & 1.487 & 1.566 & 0.750 & 0.676 & 0.617 \\
FIA        & \underline{0.759} & \underline{0.642} & \underline{0.560} & \underline{0.878} & \underline{0.792} & \underline{0.718} & \underline{0.934} & \underline{0.883} & \underline{0.833} & \underline{0.671} & \underline{0.584} & \underline{0.520} & 1.972 & 2.138 & 2.232 & \underline{0.624} & \underline{0.536} & \underline{0.479} \\
ACCENT     & 0.887 & 0.827 & 0.775 & 0.925 & 0.878 & 0.835 & 0.952 & 0.912 & 0.880 & 0.851 & 0.792 & 0.744 & 0.910 & 0.975 & 1.016 & 0.844 & 0.788 & 0.739 \\
LXR        & 0.878 & 0.810 & 0.760 & 0.921 & 0.867 & 0.830 & 0.978 & 0.905 & 0.877 & 0.867 & 0.806 & 0.760 & 0.882 & 0.965 & 1.000 & 0.805 & 0.744 & 0.697 \\
PI       & 0.769 & 0.649 & 0.571 & 0.883 & 0.799 & 0.723 & 0.936 & 0.888 & 0.835 & 0.675 & 0.588 & 0.523 & 1.979 & \underline{2.159} & \underline{2.249} & 0.631 & 0.543 & 0.484 \\
SPINRec    & \textbf{0.740} & \textbf{0.603} & \textbf{0.498} & \textbf{0.870} & \textbf{0.767} & \textbf{0.686} & \textbf{0.930} & \textbf{0.871} & \textbf{0.813} & \textbf{0.665} & \textbf{0.573} & \textbf{0.505} & \textbf{2.348} & \textbf{2.539} & \textbf{2.610} & \textbf{0.590} & \textbf{0.500} & \textbf{0.440} \\
\hline
\multicolumn{19}{|c|}{\textbf{NCF}} \\
\hline
Cosine     & 0.796 & 0.710 & 0.639 & 0.843 & 0.762 & 0.694 & 0.879 & 0.814 & 0.758 & 0.911 & 0.878 & 0.851 & 0.738 & 0.786 & 0.819 & 0.734 & 0.657 & 0.599 \\
SHAP       & 0.871 & 0.810 & 0.755 & 0.889 & 0.830 & 0.779 & 0.910 & 0.863 & 0.819 & 0.947 & 0.923 & 0.901 & 0.650 & 0.688 & 0.719 & 0.828 & 0.772 & 0.726 \\
DeepSHAP   & 0.705 & 0.583 & 0.489 & 0.767 & 0.650 & \underline{0.551} & \underline{0.817} & 0.713 & 0.636 & \underline{0.877} & \underline{0.833} & \underline{0.796} & \underline{0.793} & \underline{0.853} & \underline{0.892} & 0.655 & 0.559 & 0.481 \\
LIME       & 0.718 & 0.607 & 0.514 & 0.784 & 0.670 & 0.585 & 0.833 & 0.733& 0.658 & 0.882 & 0.840 & 0.805 & 0.782 & 0.839 & 0.877 & 0.673 & 0.580 & 0.512 \\
LIRE       & 0.826 & 0.747 & 0.669 & 0.866 & 0.789 & 0.721 & 0.899 & 0.843 & 0.782 & 0.919 & 0.887 & 0.857 & 0.728 & 0.780 & 0.820 & 0.768 & 0.696 & 0.635 \\
FIA        & \underline{0.700} & \underline{0.579} & \underline{0.484} & \underline{0.765} & \underline{0.647} & \underline{0.551} & \underline{0.817} & \underline{0.710} & \underline{0.634} & \textbf{0.876} & \underline{0.833} & \underline{0.796} & \underline{0.793} & \underline{0.853} & \underline{0.892} & \underline{0.652} & \underline{0.556} & \underline{0.481} \\
ACCENT     & 0.766 & 0.666 & 0.579 & 0.806 & 0.711 & 0.628 & 0.849 & 0.763 & 0.695 & 0.887 & 0.846 & 0.813 & 0.771 & 0.827 & 0.866 & 0.730 & 0.640 & 0.576 \\
LXR        & 0.735 & 0.628 & 0.526 & 0.786 & 0.675 & 0.590 & 0.833 & 0.738 & 0.662 & 0.889 & 0.845 & 0.808 & 0.769 & 0.832 & 0.874 & 0.689 & 0.594 & 0.519 \\
PI       & 0.704 & 0.585 & 0.489 & 0.768 & 0.653 & 0.554 & 0.819 & 0.715 & 0.636 & \underline{0.877} & \underline{0.833} & \underline{0.796} & 0.792 & 0.852 & 0.891 & 0.657 & 0.562 & 0.486 \\
SPINRec    & \textbf{0.686} & \textbf{0.558} & \textbf{0.455} & \textbf{0.757} & \textbf{0.631} & \textbf{0.535} & \textbf{0.809} & \textbf{0.703} & \textbf{0.625} & \textbf{0.876} & \textbf{0.832} &\textbf{0.795} & \textbf{0.796} & \textbf{0.856} & \textbf{0.895} & \textbf{0.621} & \textbf{0.522} & \textbf{0.448} \\
\hline
\end{tabular}
}
\end{table*}

\section{Acknowledgment}
This work was supported by the Ministry of Innovation, Science \& Technology, Israel.

\bibliography{99_references,ref}

\begin{thebibliography}{64}
\providecommand{\natexlab}[1]{#1}

\bibitem[{Abdollahi and Nasraoui(2016)}]{abdollahi2016explainable}
Abdollahi, B.; and Nasraoui, O. 2016.
\newblock Explainable matrix factorization for collaborative filtering.
\newblock In \emph{Proceedings of the 25th International Conference Companion on World Wide Web}, 5--6.

\bibitem[{Abdollahi and Nasraoui(2017)}]{abdollahi2017using}
Abdollahi, B.; and Nasraoui, O. 2017.
\newblock Using explainability for constrained matrix factorization.
\newblock In \emph{Proceedings of the eleventh ACM conference on recommender systems}, 79--83.

\bibitem[{Agarwal et~al.(2022)Agarwal, Krishna, Saxena, Pawelczyk, Johnson, Puri, Zitnik, and Lakkaraju}]{agarwal2022openxai}
Agarwal, C.; Krishna, S.; Saxena, E.; Pawelczyk, M.; Johnson, N.; Puri, I.; Zitnik, M.; and Lakkaraju, H. 2022.
\newblock Openxai: Towards a transparent evaluation of model explanations.
\newblock \emph{Advances in Neural Information Processing Systems}, 35: 15784--15799.

\bibitem[{Baklanov et~al.(2025)Baklanov, Bogina, Elisha, Schein, Allerhand, Barkan, and Koenigstein}]{mikhail_metrics}
Baklanov, M.; Bogina, V.; Elisha, Y.; Schein, Y.; Allerhand, L.; Barkan, O.; and Koenigstein, N. 2025.
\newblock Refining Fidelity Metrics for Explainable Recommendations.
\newblock In \emph{Proceedings of the 48th International ACM SIGIR Conference on Research and Development in Information Retrieval}, 2967--2971.

\bibitem[{Barkan et~al.(2024)Barkan, Bogina, Gurevitch, Asher, and Koenigstein}]{lxr}
Barkan, O.; Bogina, V.; Gurevitch, L.; Asher, Y.; and Koenigstein, N. 2024.
\newblock A Counterfactual Framework for Learning and Evaluating Explanations for Recommender Systems.
\newblock In \emph{Proceedings of the ACM on Web Conference 2024}, 3723--3733.

\bibitem[{Barkan et~al.(2023{\natexlab{a}})Barkan, Elisha, Asher, Eshel, and Koenigstein}]{barkan2023visual}
Barkan, O.; Elisha, Y.; Asher, Y.; Eshel, A.; and Koenigstein, N. 2023{\natexlab{a}}.
\newblock Visual Explanations via Iterated Integrated Attributions.
\newblock In \emph{IEEE/CVF International Conference on Computer Vision (ICCV)}, 2073--2084.

\bibitem[{Barkan et~al.(2023{\natexlab{b}})Barkan, Elisha, Asher, Eshel, and Koenigstein}]{Barkan_2023_ICCV}
Barkan, O.; Elisha, Y.; Asher, Y.; Eshel, A.; and Koenigstein, N. 2023{\natexlab{b}}.
\newblock Visual Explanations via Iterated Integrated Attributions.
\newblock In \emph{Proceedings of the IEEE/CVF International Conference on Computer Vision (ICCV)}, 2073--2084.

\bibitem[{Barkan et~al.(2023{\natexlab{c}})Barkan, Elisha, Weill, Asher, Eshel, and Koenigstein}]{barkan2023deep}
Barkan, O.; Elisha, Y.; Weill, J.; Asher, Y.; Eshel, A.; and Koenigstein, N. 2023{\natexlab{c}}.
\newblock Deep integrated explanations.
\newblock In \emph{Proceedings of the 32nd ACM International Conference on Information and Knowledge Management}, 57--67.

\bibitem[{Barkan et~al.(2023{\natexlab{d}})Barkan, Elisha, Weill, Asher, Eshel, and Koenigstein}]{barkan2023six}
Barkan, O.; Elisha, Y.; Weill, J.; Asher, Y.; Eshel, A.; and Koenigstein, N. 2023{\natexlab{d}}.
\newblock Stochastic Integrated Explanations for Vision Models.
\newblock In \emph{2023 IEEE International Conference on Data Mining (ICDM)}. IEEE.

\bibitem[{Barkan et~al.(2025)Barkan, Elisha, Weill, and Koenigstein}]{barkan2025bee}
Barkan, O.; Elisha, Y.; Weill, J.; and Koenigstein, N. 2025.
\newblock BEE: Metric-Adapted Explanations via Baseline Exploration-Exploitation.
\newblock In \emph{Proceedings of the AAAI Conference on Artificial Intelligence}, volume~39, 1835--1843.

\bibitem[{Barkan et~al.(2020)Barkan, Fuchs, Caciularu, and Koenigstein}]{barkan2020explainable}
Barkan, O.; Fuchs, Y.; Caciularu, A.; and Koenigstein, N. 2020.
\newblock Explainable recommendations via attentive multi-persona collaborative filtering.
\newblock In \emph{Proceedings of the 14th ACM Conference on Recommender Systems}, 468--473.

\bibitem[{Barkan et~al.(2021)Barkan, Hirsch, Katz, Caciularu, Weill, and Koenigstein}]{barkan2021cold}
Barkan, O.; Hirsch, R.; Katz, O.; Caciularu, A.; Weill, J.; and Koenigstein, N. 2021.
\newblock Cold item integration in deep hybrid recommenders via tunable stochastic gates.
\newblock In \emph{2021 IEEE International Conference on Data Mining (ICDM)}, 994--999. IEEE.

\bibitem[{Barkan, Katz, and Koenigstein(2020)}]{barkan2020neural}
Barkan, O.; Katz, O.; and Koenigstein, N. 2020.
\newblock Neural attentive multiview machines.
\newblock In \emph{ICASSP 2020-2020 IEEE International Conference on Acoustics, Speech and Signal Processing (ICASSP)}, 3357--3361. IEEE.

\bibitem[{Barkan et~al.(2019)Barkan, Koenigstein, Yogev, and Katz}]{barkan2019cb2cf}
Barkan, O.; Koenigstein, N.; Yogev, E.; and Katz, O. 2019.
\newblock CB2CF: a neural multiview content-to-collaborative filtering model for completely cold item recommendations.
\newblock In \emph{Proceedings of the 13th ACM Conference on Recommender Systems}, 228--236.

\bibitem[{Barkan et~al.(2023{\natexlab{e}})Barkan, Shaked, Fuchs, and Koenigstein}]{barkan2023modeling}
Barkan, O.; Shaked, T.; Fuchs, Y.; and Koenigstein, N. 2023{\natexlab{e}}.
\newblock Modeling users’ heterogeneous taste with diversified attentive user profiles.
\newblock \emph{User Modeling and User-Adapted Interaction}, 1--31.

\bibitem[{Brunot et~al.(2022)Brunot, Canovas, Chanson, Labroche, and Verdeaux}]{brunot2022preference}
Brunot, L.; Canovas, N.; Chanson, A.; Labroche, N.; and Verdeaux, W. 2022.
\newblock Preference-based and local post-hoc explanations for recommender systems.
\newblock \emph{Information Systems}, 108: 102021.

\bibitem[{Cheng et~al.(2019)Cheng, Shen, Huang, and Zhu}]{cheng2019incorporating}
Cheng, W.; Shen, Y.; Huang, L.; and Zhu, Y. 2019.
\newblock Incorporating interpretability into latent factor models via fast influence analysis.
\newblock In \emph{Proceedings of the 25th ACM SIGKDD International Conference on Knowledge Discovery \& Data Mining}, 885--893.

\bibitem[{Dror et~al.(2012)Dror, Koenigstein, Koren, and Weimer}]{dror2012yahoo}
Dror, G.; Koenigstein, N.; Koren, Y.; and Weimer, M. 2012.
\newblock The yahoo! music dataset and kdd-cup’11.
\newblock In \emph{Proceedings of KDD Cup 2011}, 3--18. PMLR.

\bibitem[{Elisha, Barkan, and Koenigstein(2024)}]{elisha2024probabilistic}
Elisha, Y.; Barkan, O.; and Koenigstein, N. 2024.
\newblock Probabilistic Path Integration with Mixture of Baseline Distributions.
\newblock In \emph{Proceedings of the 33rd ACM International Conference on Information and Knowledge Management}, 570--580.

\bibitem[{Enguehard(2023)}]{enguehard2023sequential}
Enguehard, J. 2023.
\newblock Sequential integrated gradients: a simple but effective method for explaining language models.
\newblock In Rogers, A.; Boyd-Graber, J.; and Okazaki, N., eds., \emph{Findings of the Association for Computational Linguistics: ACL 2023}, 7555--7565. Toronto, Canada: Association for Computational Linguistics.

\bibitem[{Erion et~al.(2021)Erion, Janizek, Sturmfels, Lundberg, and Lee}]{erion2021improving}
Erion, G.; Janizek, J.~D.; Sturmfels, P.; Lundberg, S.~M.; and Lee, S.-I. 2021.
\newblock Improving performance of deep learning models with axiomatic attribution priors and expected gradients.
\newblock \emph{Nature machine intelligence}, 3(7): 620--631.

\bibitem[{Fan et~al.(2022)Fan, Zhao, Chen, Su, Gao, Wang, Liu, Wang, Xu, Chen et~al.}]{fan2022comprehensive}
Fan, W.; Zhao, X.; Chen, X.; Su, J.; Gao, J.; Wang, L.; Liu, Q.; Wang, Y.; Xu, H.; Chen, L.; et~al. 2022.
\newblock A comprehensive survey on trustworthy recommender systems.
\newblock \emph{arXiv preprint arXiv:2209.10117}.

\bibitem[{Gaiger et~al.(2023)Gaiger, Barkan, Tsipory-Samuel, and Koenigstein}]{ai2vpp23}
Gaiger, K.; Barkan, O.; Tsipory-Samuel, S.; and Koenigstein, N. 2023.
\newblock Not All Memories Created Equal: Dynamic User Representations for Collaborative Filtering.
\newblock \emph{IEEE Access}, 1--1.

\bibitem[{Gurevitch et~al.(2025)Gurevitch, Bogina, Barkan, Schein, Elisha, and Koenigstein}]{gurevitchlxr}
Gurevitch, L.; Bogina, V.; Barkan, O.; Schein, Y.; Elisha, Y.; and Koenigstein, N. 2025.
\newblock LXR: Learning to eXplain Recommendations.
\newblock \emph{ACM Transactions on Recommender Systems}.

\bibitem[{Harper and Konstan(2015)}]{harper2015movielens}
Harper, F.~M.; and Konstan, J.~A. 2015.
\newblock The movielens datasets: History and context.
\newblock \emph{Acm transactions on interactive intelligent systems (tiis)}, 5(4): 1--19.

\bibitem[{Haug et~al.(2021)Haug, Z{\"u}rn, El-Jiz, and Kasneci}]{haug2021baselines}
Haug, J.; Z{\"u}rn, S.; El-Jiz, P.; and Kasneci, G. 2021.
\newblock On baselines for local feature attributions.
\newblock \emph{arXiv preprint arXiv:2101.00905}.

\bibitem[{He et~al.(2020)He, Deng, Wang, Li, Zhang, and Wang}]{he2020lightgcn}
He, X.; Deng, K.; Wang, X.; Li, Y.; Zhang, Y.; and Wang, M. 2020.
\newblock Lightgcn: Simplifying and powering graph convolution network for recommendation.
\newblock In \emph{Proceedings of the 43rd International ACM SIGIR conference on research and development in Information Retrieval}, 639--648.

\bibitem[{He et~al.(2017)He, Liao, Zhang, Nie, Hu, and Chua}]{he2017neural}
He, X.; Liao, L.; Zhang, H.; Nie, L.; Hu, X.; and Chua, T.-S. 2017.
\newblock Neural collaborative filtering.
\newblock In \emph{Proceedings of the 26th international conference on world wide web}, 173--182.

\bibitem[{Kang and McAuley(2018)}]{kang2018self}
Kang, W.-C.; and McAuley, J. 2018.
\newblock Self-attentive sequential recommendation.
\newblock In \emph{2018 IEEE international conference on data mining (ICDM)}, 197--206. IEEE.

\bibitem[{Kapishnikov et~al.(2021)Kapishnikov, Venugopalan, Avci, Wedin, Terry, and Bolukbasi}]{kapishnikov2021guided}
Kapishnikov, A.; Venugopalan, S.; Avci, B.; Wedin, B.; Terry, M.; and Bolukbasi, T. 2021.
\newblock Guided integrated gradients: An adaptive path method for removing noise.
\newblock In \emph{Proceedings of the IEEE/CVF conference on computer vision and pattern recognition}, 5050--5058.

\bibitem[{Katz et~al.(2022)Katz, Barkan, Koenigstein, and Zabari}]{katz2022learning}
Katz, O.; Barkan, O.; Koenigstein, N.; and Zabari, N. 2022.
\newblock Learning to ride a buy-cycle: A hyper-convolutional model for next basket repurchase recommendation.
\newblock In \emph{Proceedings of the 16th ACM Conference on Recommender Systems}, 316--326.

\bibitem[{Koenigstein(2025)}]{koenigstein2025without}
Koenigstein, N. 2025.
\newblock Without Fidelity, Explanations Are Just Stories: Rethinking Evaluation in Explainable Recommender Systems.
\newblock \emph{SSRN (September 26, 2025)}.

\bibitem[{Koh and Liang(2017)}]{koh2017understanding}
Koh, P.~W.; and Liang, P. 2017.
\newblock Understanding black-box predictions via influence functions.
\newblock In \emph{International conference on machine learning}, 1885--1894. PMLR.

\bibitem[{Koren, Bell, and Volinsky(2009)}]{koren2009matrix}
Koren, Y.; Bell, R.; and Volinsky, C. 2009.
\newblock Matrix factorization techniques for recommender systems.
\newblock \emph{Computer}, 42(8): 30--37.

\bibitem[{Kunkel et~al.(2019)Kunkel, Donkers, Michael, Barbu, and Ziegler}]{kunkel2019let}
Kunkel, J.; Donkers, T.; Michael, L.; Barbu, C.-M.; and Ziegler, J. 2019.
\newblock Let me explain: Impact of personal and impersonal explanations on trust in recommender systems.
\newblock In \emph{Proceedings of the 2019 CHI conference on human factors in computing systems}, 1--12.

\bibitem[{Li, Zhang, and Chen(2021)}]{li2021personalized}
Li, L.; Zhang, Y.; and Chen, L. 2021.
\newblock Personalized Transformer for Explainable Recommendation.
\newblock In \emph{Proceedings of the 59th Annual Meeting of the Association for Computational Linguistics and the 11th International Joint Conference on Natural Language Processing (Volume 1: Long Papers)}, 4947--4957.

\bibitem[{Liang et~al.(2018)Liang, Krishnan, Hoffman, and Jebara}]{liang2018variational}
Liang, D.; Krishnan, R.~G.; Hoffman, M.~D.; and Jebara, T. 2018.
\newblock Variational autoencoders for collaborative filtering.
\newblock In \emph{Proceedings of the 2018 world wide web conference}, 689--698.

\bibitem[{Lundberg and Lee(2017{\natexlab{a}})}]{deep_shap}
Lundberg, S.; and Lee, S.-I. 2017{\natexlab{a}}.
\newblock A unified approach to interpreting model predictions.
\newblock \emph{Advances in Neural Information Processing Systems}, 4765–--4774.

\bibitem[{Lundberg and Lee(2017{\natexlab{b}})}]{shap_original}
Lundberg, S.~M.; and Lee, S.-I. 2017{\natexlab{b}}.
\newblock A unified approach to interpreting model predictions.
\newblock \emph{Advances in neural information processing systems}, 30.

\bibitem[{Melchiorre et~al.(2022)Melchiorre, Rekabsaz, Ganh{\"o}r, and Schedl}]{melchiorre2022protomf}
Melchiorre, A.~B.; Rekabsaz, N.; Ganh{\"o}r, C.; and Schedl, M. 2022.
\newblock ProtoMF: Prototype-based Matrix Factorization for Effective and Explainable Recommendations.
\newblock In \emph{Sixteenth ACM Conference on Recommender Systems (RecSys '22)}, 11. Seattle, WA, USA: ACM.

\bibitem[{Mohammadi et~al.(2025)Mohammadi, Peintner, M{\"u}ller, and Zangerle}]{Mohammadi2025beyond}
Mohammadi, A.~R.; Peintner, A.; M{\"u}ller, M.; and Zangerle, E. 2025.
\newblock Beyond Top-1: Addressing Inconsistencies in Evaluating Counterfactual Explanations for Recommender Systems.
\newblock In \emph{Proceedings of the Nineteenth ACM Conference on Recommender Systems}, 515--520.

\bibitem[{N{\'o}brega and Marinho(2019)}]{nobrega2019towards}
N{\'o}brega, C.; and Marinho, L. 2019.
\newblock Towards explaining recommendations through local surrogate models.
\newblock In \emph{Proceedings of the 34th ACM/SIGAPP Symposium on Applied Computing}, 1671--1678.

\bibitem[{Rendle et~al.(2022)Rendle, Krichene, Zhang, and Koren}]{rendle2022revisiting}
Rendle, S.; Krichene, W.; Zhang, L.; and Koren, Y. 2022.
\newblock Revisiting the performance of ials on item recommendation benchmarks.
\newblock In \emph{Proceedings of the 16th ACM Conference on Recommender Systems}, 427--435.

\bibitem[{Ribeiro, Singh, and Guestrin(2016)}]{lime_original}
Ribeiro, M.~T.; Singh, S.; and Guestrin, C. 2016.
\newblock ``Why should i trust you?'' Explaining the predictions of any classifier.
\newblock In \emph{Proceedings of the 22nd ACM SIGKDD international conference on knowledge discovery and data mining}, 1135--1144.

\bibitem[{Samek et~al.(2016)Samek, Binder, Montavon, Lapuschkin, and M{\"u}ller}]{samek2016evaluating}
Samek, W.; Binder, A.; Montavon, G.; Lapuschkin, S.; and M{\"u}ller, K.-R. 2016.
\newblock Evaluating the visualization of what a deep neural network has learned.
\newblock \emph{IEEE transactions on neural networks and learning systems}, 28(11): 2660--2673.

\bibitem[{Sanyal and Ren(2021)}]{sanyal2021discretized}
Sanyal, S.; and Ren, X. 2021.
\newblock Discretized Integrated Gradients for Explaining Language Models.
\newblock In Moens, M.-F.; Huang, X.; Specia, L.; and Yih, S. W.-t., eds., \emph{Proceedings of the 2021 Conference on Empirical Methods in Natural Language Processing}, 10285--10299. Online and Punta Cana, Dominican Republic: Association for Computational Linguistics.

\bibitem[{Shenbin et~al.(2020)Shenbin, Alekseev, Tutubalina, Malykh, and Nikolenko}]{shenbin2020recvae}
Shenbin, I.; Alekseev, A.; Tutubalina, E.; Malykh, V.; and Nikolenko, S.~I. 2020.
\newblock Recvae: A new variational autoencoder for top-n recommendations with implicit feedback.
\newblock In \emph{Proceedings of the 13th international conference on web search and data mining}, 528--536.

\bibitem[{Shrikumar, Greenside, and Kundaje(2017)}]{shrikumar2017learning}
Shrikumar, A.; Greenside, P.; and Kundaje, A. 2017.
\newblock Learning important features through propagating activation differences.
\newblock In \emph{International conference on machine learning}, 3145--3153. PMlR.

\bibitem[{Sikdar, Bhattacharya, and Heese(2021)}]{sikdar2021integrated}
Sikdar, S.; Bhattacharya, P.; and Heese, K. 2021.
\newblock Integrated directional gradients: Feature interaction attribution for neural NLP models.
\newblock In \emph{Proceedings of the 59th Annual Meeting of the Association for Computational Linguistics and the 11th International Joint Conference on Natural Language Processing (Volume 1: Long Papers)}, 865--878.

\bibitem[{Singh et~al.(2020)Singh, Maurya, Tripathi, Narula, and Srivastav}]{singh2020movie}
Singh, R.~H.; Maurya, S.; Tripathi, T.; Narula, T.; and Srivastav, G. 2020.
\newblock Movie recommendation system using cosine similarity and KNN.
\newblock \emph{International Journal of Engineering and Advanced Technology}, 9(5): 556--559.

\bibitem[{Sturmfels, Lundberg, and Lee(2020)}]{sturmfels2020visualizing}
Sturmfels, P.; Lundberg, S.; and Lee, S.-I. 2020.
\newblock Visualizing the Impact of Feature Attribution Baselines.
\newblock \emph{Distill}.
\newblock Https://distill.pub/2020/attribution-baselines.

\bibitem[{Sugahara and Okamoto(2024)}]{sugahara2024hierarchical}
Sugahara, K.; and Okamoto, K. 2024.
\newblock Hierarchical matrix factorization for interpretable collaborative filtering.
\newblock \emph{Pattern Recognition Letters}, 180: 99--106.

\bibitem[{Sundararajan, Taly, and Yan(2017)}]{sundararajan2017axiomatic}
Sundararajan, M.; Taly, A.; and Yan, Q. 2017.
\newblock Axiomatic attribution for deep networks.
\newblock In \emph{International conference on machine learning}, 3319--3328. PMLR.

\bibitem[{Tintarev(2025)}]{tintarevmeasuring}
Tintarev, N. 2025.
\newblock Measuring Explanation Quality--A Path Forward.
\newblock In \emph{ECAI 2025}, 22--29. IOS Press.

\bibitem[{Tintarev and Masthoff(2015)}]{tintarev2015explaining}
Tintarev, N.; and Masthoff, J. 2015.
\newblock Explaining recommendations: Design and evaluation.
\newblock In \emph{Recommender systems handbook}, 353--382. Springer.

\bibitem[{Tintarev and Masthoff(2022)}]{tintarev2022beyond}
Tintarev, N.; and Masthoff, J. 2022.
\newblock Beyond explaining single item recommendations.
\newblock \emph{Recommender Systems Handbook}, 711--756.

\bibitem[{Tran, Ghazimatin, and Saha~Roy(2021)}]{tran2021counterfactual}
Tran, K.~H.; Ghazimatin, A.; and Saha~Roy, R. 2021.
\newblock Counterfactual explanations for neural recommenders.
\newblock In \emph{Proceedings of the 44th International ACM SIGIR Conference on Research and Development in Information Retrieval}, 1627--1631.

\bibitem[{Varasteh et~al.(2024)Varasteh, McKinnie, Aird, Acu{\~n}a, and Burke}]{varasteh2024comparative}
Varasteh, M.; McKinnie, E.; Aird, A.; Acu{\~n}a, D.; and Burke, R. 2024.
\newblock Comparative Explanations for Recommendation: Research Directions.
\newblock \emph{IntRS’24: Joint Workshop on Interfaces and Human Decision Making for Recommender Systems}.

\bibitem[{Vig, Sen, and Riedl(2009)}]{vig2009tagsplanations}
Vig, J.; Sen, S.; and Riedl, J. 2009.
\newblock Tagsplanations: explaining recommendations using tags.
\newblock In \emph{Proceedings of the 14th international conference on Intelligent user interfaces}, 47--56.

\bibitem[{Wang et~al.(2018)Wang, Wang, Jia, and Yin}]{wang2018explainable}
Wang, N.; Wang, H.; Jia, Y.; and Yin, Y. 2018.
\newblock Explainable recommendation via multi-task learning in opinionated text data.
\newblock In \emph{The 41st International ACM SIGIR Conference on Research \& Development in Information Retrieval}, 165--174.

\bibitem[{Xu, Venugopalan, and Sundararajan(2020)}]{xu2020attribution}
Xu, S.; Venugopalan, S.; and Sundararajan, M. 2020.
\newblock Attribution in scale and space.
\newblock In \emph{Proceedings of the IEEE/CVF Conference on Computer Vision and Pattern Recognition}, 9680--9689.

\bibitem[{Zhang, Chen et~al.(2020)}]{zhang2020explainable}
Zhang, Y.; Chen, X.; et~al. 2020.
\newblock Explainable recommendation: A survey and new perspectives.
\newblock \emph{Foundations and Trends{\textregistered} in Information Retrieval}, 14(1): 1--101.

\bibitem[{Zhang et~al.(2014)Zhang, Lai, Zhang, Zhang, Liu, and Ma}]{zhang2014explicit}
Zhang, Y.; Lai, K.; Zhang, W.; Zhang, Y.; Liu, Y.; and Ma, S. 2014.
\newblock Explicit factor models for explainable recommendation based on phrase-level sentiment analysis.
\newblock In \emph{Proceedings of the 37th international ACM SIGIR conference on Research \& development in information retrieval}.

\bibitem[{Zhong and Negre(2022)}]{shap4rec}
Zhong, J.; and Negre, E. 2022.
\newblock Shap-enhanced counterfactual explanations for recommendations.
\newblock In \emph{Proceedings of the 37th ACM/SIGAPP Symposium on Applied Computing}, 1365--1372.

\end{thebibliography}

\end{document}